\documentclass[aps,pra,amssymb,amsmath,nobibnotes,nobibnotes,reprint,superscriptaddress,showpacs,showkeys]{revtex4-1}
\usepackage[colorlinks, linkcolor=blue, anchorcolor=blue, citecolor=blue]{hyperref}
\usepackage{bm, amsfonts, mathrsfs}
\usepackage{verbatim, psfrag, times, CJK}
\usepackage{graphics, graphicx, color, epsfig}
\usepackage{float}
\usepackage{flafter,natbib}
\usepackage{indentfirst}

\begin{document}

\newcommand{\ket}[1]{| #1 \rangle}
\newcommand{\bra}[1]{\langle #1 |}

\title{Establishment of correlated states in a quantum dot interacting with an acoustic phonon reservoir}
\author{Hui Huang}
\affiliation{Department of Physics, Huazhong Normal University, Wuhan 430079, P. R. China}
\author{Gao-xiang Li}
\email{gaox@phy.ccnu.edu.cn}
\affiliation{Department of Physics, Huazhong Normal University, Wuhan 430079, P. R. China}
\author{Wen-ju Gu}
\affiliation{Department of Physics, Huazhong Normal University, Wuhan 430079, P. R. China}
\author{Zbigniew Ficek}
\affiliation{The National Center for Mathematics and Physics, KACST, P.O. Box 6086, Riyadh 11442, Saudi Arabia}

\begin{abstract}
We investigate the effects of a low frequency (acoustic) phonon bath on the dynamics of a quantum dot modelled as a cascade three-level system. We show that the phonon bath appears to the upper transition of the cascade system as a broadband reservoir of inverted rather than conventional harmonic oscillators. The action of the annihilation and creation operators of the inverted oscillator are interchanges relative to those of the usual harmonic oscillator that it serves as a linear amplifier to the system, and thereby gives rise to unusual features in the dynamics of the quantum dot. We find that the phonon bath, although being in a thermal state, affects the quantum dot as a correlated-type reservoir which results in the decay of the system to a correlated two-photon state with the population distribution no longer obeying a Boltzmann distribution. It is particularly interesting that even for a zero temperature phonon reservoir the steady state is a correlated state which under appropriate conditions on the Rabi frequencies and the damping rates can reduce to a strongly correlated pure state. It is shown that the two-photon correlations result in a significant squeezing and strong two-photon correlations in the radiation field emitted by the quantum dot. The presence of the correlations in the system is manifest in the presence of quantum beats in the time evolution of the populations and the radiation intensity. The effect of the ordinary spontaneous emission on the features induced by the phonon bath is also discussed.
\end{abstract}

\pacs{42.50.Ct, 42.50.Lc, 78.67.Hc}

\maketitle

\section{Introduction}

The interaction of a quantum dot (QD) with a phonon reservoir has been studied extensively, and a number of interesting effects has been predicted. Most of these studies has considered the QD consisting of a single electron-hole pair, a two-state system, and many features characteristic of two-state system, such as the Mollow triplet, the Autler-Townes doublet and vacuum Rabi splitting have been predicted and experimentally observed~\cite{fm09,vz09,xs07,ua11,rh11,rh12}.

Recently, interest has arisen in the problem of sensitivity of a QD to the nature of a phonon bath to which it is coupled. The phonon bath serves as a low frequency reservoir and, as a result, the radiative properties of the QD change. Various problems related to the temperature of the phonon bath have been studied including phonon mediated excitation transfer~\cite{rm08,mb12}, damping of Rabi oscillations~\cite{rg10,mt13,fw03,vc07,vc11,md11}, modifications of the fluorescence spectrum~\cite{mp12,mn13,wh13}, and the creation of a steady-state population inversion between the bare states of a two-level QD system~\cite{wu12,hc13,dm13}. The population inversion gives the possibility to achieve lasing in the two-level system~\cite{zh13,zh14}.

In the case of a QD, the most important damping mechanism is a decay associated with the coupling of the QD to a finite temperature phonon bath. In general, an excited QD is expected to decay to a mixed state characteristic of conventional thermodynamic equilibrium with the population distribution that obeys a Boltzmann distribution. The decay process can be modified and the nature of the equilibrium state could be different if the QD decays in a correlated reservoir, such as a squeezed vacuum~\cite{ep89,ap90,zf91,fd91,bk91,mz03,lg04,gl13}.

The energy structure of a QD does not have to be confined to two levels only, that is to a single exciton state.  It can be extended to include a biexciton state, which could be realized in a semiconductor QD consisting of two electron-hole pairs driven by appropriate laser pulses with the help of phonons~\cite{gbl13,gb13,da13}. In this paper we examine the effect of a phonon thermal bath on the dynamics of a single QD modelled as a three-level system of the cascade configuration. In addition to the phonon bath, the QD is driven by a single laser field which couples to both transitions of the cascade system. Using the master equation approach, we calculate the steady-state populations of the energy levels, transient behaviours of the populations and the radiation intensity, nonclassical features such as squeezing and antibunching. We show that the phonon bath couples to the upper transition of the cascade system as a broadband reservoir of {\it inverted} harmonic oscillators, which serves as a linear amplifier to the system~\cite{wk86}. This causes the phonon bath to behave as a correlated rather than the conventional thermal reservoir, which may lead to many interesting features such as one and two-photon population inversions, squeezing, super-bunching and anti-bunching. Even though the interaction of the QD with the phonon reservoir is a dissipative process, we demonstrate that the reservoir can turn the system to decay to a strongly correlated state. Depending on the temperature of the reservoir, the correlated steady state of the QD can be a pure state which reflects these correlations.

The paper is organized as follows. In Sec.~\ref{sec2}, the effective Hamiltonian of the quantum dot interacting with a phonon bath is derived. We then apply the Hamiltonian to derive the master equation for the reduced density operator describing the properties of the quantum dot interacting with the phonon bath. We also include a possible coupling of the quantum dot to the ordinary vacuum modes that leads to the radiative spontaneous emission. The steady-state population distributions are presented in Sec.~\ref{sec3}, along with a discussion of one and two-photon inversions. Squeezing and the second-order coherence function are calculated in Sec.~\ref{sec4}. A detailed discussion of the transient behaviour of the system is presented in Sec.~\ref{sec5}. We find that the temporal behaviour of the population distribution and the radiation intensity may differ dramatically for different initial conditions. The effect of the ordinary spontaneous emission on the features induced by the phonon bath is discussed in Sec.~\ref{sec6}. Finally, in Sec.~\ref{sec7}, we summarize our results.

\section{Effective Hamiltonian of the system}\label{sec2}

\begin{figure}[h]
\begin{center}
\begin{tabular}{c}
\includegraphics[width=0.4\columnwidth]{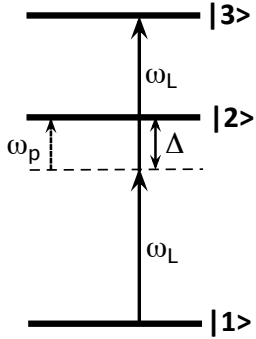}
\end{tabular}
\end{center}
\caption{Energy level scheme of the three-level quantum dot under a laser excitation of frequency $\omega_{L}$. The laser is on two-photon resonance with the $\ket 3\leftrightarrow\ket 1$ transition frequency, detuned from the transition frequency $\omega_{21}$ by $\Delta$. }
\label{fig1}
\end{figure}

We consider a quantum dot modelled as a non-degenerate three-level cascade system with the ground level~$\ket{1}$, the intermediate level $\ket 2$ and the upper level $\ket 3$ of energies $E_{3}>E_{2}>E_{1}$. The levels are separated by energies $E_{2}-E_{1}=\hbar(\omega_{2}-\omega_{1})=\hbar\omega_{21}$ and $E_{3}-E_{2}=\hbar(\omega_{3}-\omega_{2})=\hbar\omega_{32}$. In practical terms, the level $\ket 1$ corresponds to the electronic ground state of the QD, the level $\ket 2$ to the single-exciton state, and $\ket 3$ to the biexciton state. The system is driven by a single-mode laser field of frequency~$\omega_{L}$, which is resonant with a two-photon transition from $\ket 3$ to $\ket 1$, $(E_{3}-E_{1}=2\hbar\omega_{L})$, as shown in Fig.~\ref{fig1}. Because of the unequal spacing of the energy levels, the laser is not resonant with the one-photon transitions. In addition, the system interacts with modes of a low frequency (acoustic) phonon reservoir, which we assume are in a thermal state with the average occupation phonon number of a mode $p$, $\bar{n}_{p}=[\exp(\hbar\omega_{p}/k_{B}T)-1]^{-1}$, where $k_{B}$ is the Boltzmann constant and $T$ is the temperature of the reservoir. The density of the phonon modes is considered to be large only in the frequency range close to the detuning $\Delta=\omega_{21}-\omega_{L}$ of the laser frequency from the transition frequency $\omega_{21}$.

The total Hamiltonian for this system may be written in the form $(\hbar\equiv 1)$
\begin{equation}
H = H_{0} +H_{1} +H_{2} ,\label{H}
\end{equation}
where $H_{0}$ is the free Hamiltonian of the QD and the phonon reservoir
\begin{equation}
H_{0} = \Delta A_{22} + \sum_{p}\omega_{p}b_{p}^\dag b_{p} ,
\end{equation}
$H_{1}$ is the interaction Hamiltonian between the QD and the laser field mode
\begin{equation}
H_{1} = \chi_{1}\left(A_{21}+A_{12}\right) +\chi_{3}\left(A_{32} +A_{23}\right) ,\label{H1}
\end{equation}
and $H_{2}$ is the interaction Hamiltonian between the QD and the phonon reservoir
\begin{equation}
H_{2} = \sum_{p}\left(g_{3p}A_{33}-g_{1p}A_{11}\right)\left(b_{p}^\dag + b_{p}\right) .\label{H2}
\end{equation}
Here, $A_{ij}=\ket{i}\bra{j}$ are the usual atomic operators representing populations $(i=j)$ of the energy levels of the~QD and coherences $(i\neq j)$ between them, $b_{p}^{\dag}$ and $b_{p}$ are the creation and annihilation operators of mode $p$ of the phonon reservoir, $\Delta=\omega_{21}-\omega_{L}$ is the one-photon detuning of the laser frequency $\omega_{L}$ from the transition frequency $\omega_{21}$, and $\chi_1$ and $\chi_3$ are the Rabi frequencies associated with the laser field driving $\ket{1}\leftrightarrow \ket{2}$ and $\ket{2}\leftrightarrow \ket{3}$ transitions, respectively. The parameters $g_{1p}$ and $g_{3p}$ are the coupling strengths of the mode $p$ of the phonon reservoir to the one-photon transitions $\ket{2}\leftrightarrow \ket{1}$ and $\ket{3}\leftrightarrow \ket{2}$, respectively.

We make the unitary transformation of the Hamiltonian
\begin{equation}
\tilde{H} = {\rm e}^{iS}H{\rm e}^{-iS} ,\label{e5}
\end{equation}
with
\begin{equation}
S= \sum_{p}\frac{-i}{\omega_{p}}\left(g_{3p}A_{33}-g_{1p}A_{11}\right)\left(b_{p}^\dag - b_{p}\right) ,
\end{equation}
and find
\begin{equation}
\tilde{H} = \tilde{H}_{0} +\tilde{H}_{F} +\tilde{V} ,\label{e7}
\end{equation}
in which
\begin{align}
\tilde{H}_{0} &= \Delta A_{22} +\sum_{i=1,3}\Omega_{i} \left(A_{2i} +A_{i2}\right) ,\nonumber\\
\tilde{H}_{F} &= \sum\limits_p\omega_p b^\dag_p b_{p} ,\label{e8}
\end{align}
and
\begin{align}
\tilde{V} &=\sum_{p} \frac{g_{1p}\Omega_{1}}{\omega_{p}}\left(b_{p}^{\dag} - b_{p}\right)\left(A_{21}-A_{12}\right) \nonumber\\
&+ \sum_{p}\frac{g_{3p}\Omega_{3}}{\omega_{p}}\left(b_{p}^{\dag} - b_{p}\right)\left(A_{32}-A_{23}\right) ,\label{e9}
\end{align}
where $\Omega_{i}=\langle B_{i}\rangle \chi_{i}$, with
\begin{equation}
\langle B_i\rangle=\exp\left[-\frac{1}{2}\sum\limits_p (g_{ip}/\omega_p)^2(2\bar{n}_p+1)\right] \label{e10}
\end{equation}
is an effective Rabi frequency of the laser field.
In the derivation of Eq.~(\ref{e7}) we have performed a Born approximation which corresponds to a first-order term in a systematic expansion of~Eq.~(\ref{e5}) in $g_{ip}$, which means that we assumed that the interaction between the QD and the phonon reservoir is not very strong so that the one-phonon transitions play a dominant role as compared with the multiphonon transitions. 

The interaction Hamiltonian $\tilde{V}$ contains the rotating and counter-rotating terms. We may transform the Hamiltonian~(\ref{e7}) to the interaction picture with the unitary transformation $U(t) =\exp[-i(\tilde{H}_{0}+\tilde{H}_{F})t]$, and find
\begin{align}
\tilde{V}(t) &=\sum\limits_{p}\left\{ b_p^{\dag}{\rm e}^{i\left(\omega_p-\Delta\right)t}\left[\frac{g_{3p}\Omega_{3}}{\omega_p}A_{32} -\frac{g_{1p}\Omega_{1}}{\omega_p}A_{12}\right]\right. \nonumber\\
&\left. +\, b_p^{\dag}{\rm e}^{i\left(\omega_{p}+\Delta\right)t}\!\left[\frac{g_{1p}\Omega_{1}}{\omega_p}A_{21} -\frac{g_{3p}\Omega_{3}}{\omega_p}A_{23}\right]\!\right\} + {\rm H.c.} \label{e11}
\end{align}
It what follows we consider the situation where $\omega_p\approx\Delta$ and the coupling strengths $g_{i}\Omega_{i}/\omega_{p}$ are much smaller than $\omega_{p}$, $(g_{i}\Omega_{i}/\omega_{p})\ll\omega_{p}$. This prompts us to make the rotating-wave approximation in which we ignore rapidly oscillating terms at frequency $2\omega_{p}$, and obtain
\begin{align}
\tilde{V}(t)&\approx\sum\limits_p b_p^{\dag}e^{i\left(\omega_p-\Delta\right)t}\left(\tilde{g}_{3p}A_{32} -\tilde{g}_{1p}A_{12}\right) + {\rm H.c.} ,\label{e12}
\end{align}
where $\tilde{g}_{ip}= g_{ip}\Omega_{i}/\Delta$ is an effective coupling strength of the phonon reservoir to the $i$th transition of the QD. Thus, we have arrived to the effective Hamiltonian with parameters that can be controlled through the laser frequency and amplitude and where the characteristic frequency scales are no longer those of optical frequencies but those associated with detunings and Rabi frequencies.

The properties of entire system of the QD interacting with a phonon bath are described by the density operator $\rho_{T}$. The reduced density operator $\rho$ describing the properties of the QD is obtained by taking the trace of $\rho_{T}$ over the space of the phonon modes, $\rho={\rm Tr}_{p}\rho_{T}$. Assuming that the phonon modes provide a broadband reservoir we may perform the Markov approximation, as is usually done when dealing with spontaneous emission processes. Actually the non-Markovian effects are likely to affect the QD dynamics only at short times; the difference between non-Markovian and Markovian dynamics diminishes or vanishes at longer time scales~\cite{mn10}. It is reported that for a QD transition driven by a detuned continuous-wave laser by mediation of acoustic phonons, the Markov approximation is found to be an excellent approximation to describe the effect of the interaction between the QD and the phonon reservoir on the properties of the QD~\cite{rh11,rh12,wu12,ul13}. In addition, taking into account radiation damping through the ordinary spontaneous emission, we derive the master equation for the reduced density operator~$\rho$~as
\begin{eqnarray}
\frac{\partial}{\partial t}{\rho} = -i[\tilde{H}_{0},\rho] +{\cal L}_{p}\rho +{\cal L}_{s2}\rho +{\cal L}_{s3}\rho ,\label{e13}
\end{eqnarray}
in which
\begin{eqnarray}
{\cal L}_{p}\rho &=& \sum_{j=1,3}(\bar{n}+1)\gamma_{j}\left\{\left[A_{j2}\rho, A_{2j}\right] +\left[A_{j2},\rho A_{2j}\right]\right\} \nonumber\\
&+&\sum_{j=1,3}\bar{n}\gamma_{j}\left\{\left[A_{2j}\rho, A_{j2}\right] +\left[A_{2j},\rho A_{j2}\right]\right\} \nonumber\\
&-&(\bar{n}+1)\gamma_{13}\left\{\left[A_{32}\rho, A_{21}\right] +\left[A_{32},\rho A_{21}\right] + {\rm H.c.}\right\}\nonumber\\
&-&\bar{n}\gamma_{13}\left\{\left[A_{21}\rho, A_{32}\right] +\left[A_{21}, \rho A_{32}\right] +{\rm H.c.}\right\} \label{e14}
\end{eqnarray}
describes phonon-bath-induced decay of the QD. The coefficients $\gamma_{1}$ and $\gamma_{3}$ are the phonon-bath-induced damping rates of the levels $\ket 2$ and $\ket 3$, respectively, and~$\gamma_{13}=\sqrt{\gamma_{1}\gamma_{3}}$ is the cross damping rate resulting from the phonon-bath-induced dissipative coupling between the transitions.

The remaining two terms ${\cal L}_{s2}\rho$ and ${\cal L}_{s3}\rho$ are of the form
\begin{align}
{\cal L}_{s2}\rho &= \Gamma_{2} \left\{\left[A_{12}\rho, A_{21}\right] +\left[A_{12},\rho A_{21}\right]\right\} ,\nonumber\\
{\cal L}_{s3}\rho &= \Gamma_{3} \left\{\left[A_{23}\rho, A_{32}\right] +\left[A_{23},\rho A_{32}\right]\right\} \label{e15}
\end{align}
and represent de-excitation of the levels $\ket 2$ and $\ket 3$ by radiative spontaneous emission with damping rates $\Gamma_{2}$ and $\Gamma_{3}$, respectively.

\begin{figure}[h]
\begin{center}
\begin{tabular}{c}
\includegraphics[width=0.55\columnwidth]{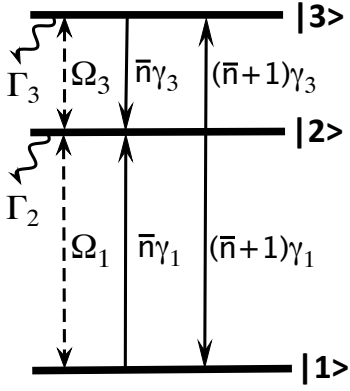}
\end{tabular}
\end{center}
\caption{Three-level cascade system interacting with a phonon bath and driven by a coherent laser field. The phonon bath couples to the transition $\ket 1\leftrightarrow \ket 2$ as an ordinary harmonic oscillator thermal bath whereas it couples to the $\ket 2\leftrightarrow \ket 3$ transition as an inverted harmonic oscillator thermal bath. In addition levels $\ket 1$ and $\ket 2$ are coupled by the driving laser field with the Rabi frequency $\Omega_{1}$ and levels $\ket 2$ and $\ket 3$ with the Rabi frequency $\Omega_{3}$. Level $\ket 3$ decays to level $\ket 2$ with rate $\Gamma_{3}$ and level $\ket 2$ decays to level $\ket 1$ with rate $\Gamma_{2}$.}
\label{fig2}
\end{figure}

There are several interesting remarks that should be made about the expression for the Liouvillean ${\cal L}_{p}\rho$, Eq.~(\ref{e14}). First of all, we may identify the first two terms with incoherent damping and incoherent pumping of the two transitions. However, a close look reveals that the phonon reservoir couples to the $\ket 1\leftrightarrow \ket 2$ and $\ket 2\leftrightarrow \ket 3$ transitions in decidedly different ways. It is easy to see that the first term corresponds to the damping of the $\ket 1\leftrightarrow \ket 2$ transition but it appears as an incoherent pumping of the $\ket 3\leftrightarrow \ket 2$ transition. Similarly, the second term corresponds to an incoherent pumping of the $\ket 1\leftrightarrow \ket 2$ transition but it appears as an incoherent damping of the $\ket 3\leftrightarrow \ket 2$ transition. In other words, the phonon bath couples to the lower transition $\ket 1\leftrightarrow \ket 2$ as an ordinary harmonic oscillator thermal bath with damping rate $(\bar{n}+1)\gamma_{1}$ and with incoherent pumping rate $\bar{n}\gamma_{1}$, whereas it couples to the upper transition $\ket 2\leftrightarrow\ket 3$ as an {\it inverted} harmonic oscillator thermal bath with damping rate $\bar{n}\gamma_{3}$ and with incoherent pumping rate $(\bar{n}+1)\gamma_{3}$. Figure~\ref{fig2} helps to explain the situation. It illustrates the action of the phonon bath on the transitions between the energy levels of the system.

What is more surprising than the coupling to the upper transition as the inverted harmonic oscillator is the presence in Eq.~(\ref{e14}) of two terms involving cross coupling between the transitions $\ket 1\leftrightarrow \ket 2$ and $\ket 2\leftrightarrow \ket 3$. This is a form of interference, although it involves transitions of ordinary and inverted harmonic oscillators. In particular, the third term in Eq.~(\ref{e14}) reflects the fact that, as the QD decays from the level $\ket 2$ to the level~$\ket 1$, it drives absorption to level $\ket 3$. Similarly, the forth term reflects the fact that, as the QD is incoherently pumped from the level~$\ket 1$ to level~$\ket 2$, it drives emission from the level $\ket 3$ to level $\ket 2$. In physical terms, the third term in Eq.~(\ref{e14}) represents the process of a de-excitation of the level $\ket 2$ simultaneously to the levels $\ket 1$ and $\ket 3$, whereas the fourth term represents the process of an excitation of the level $\ket 2$ simultaneously from the levels $\ket 1$ and $\ket 3$. We shall see that the effect of the simultaneous de-excitation of the level $\ket 2$ to levels $\ket 1$ and $\ket 3$ may result in a strong coherence between these levels.

Thus, we may conclude that the phonon bath, although being in a thermal state, affects the QD as a correlated-type reservoir. One could argue that the action of the phonon bath on the system is similar in form to the ones of a squeezed reservoir coupled to a cascade three-level system. However, the analogy is not complete. For example, the cross coupling between the transitions preserves even in the case of zero temperature phonon bath, where $\bar{n}=0$. This is in strike contrast to the squeezed reservoir that requires $\bar{n}\neq 0$ for the cross coupling to be present.
Although the analogy is not complete, we shall show that various results predicted for the cascade system interacting with a phonon bath are analogous to that obtained in the cascade system interacting with a squeezed reservoir~\cite{fd91,bk91}.

\section {Coherent superposition states}

The presence of the cross coupling (interference) terms in Eq.~(\ref{e14}) indicates that the phonon bath induces a direct coupling between the states $\ket 1$ and $\ket 3$, which may result in a coherence between these states. When the coherence is present, the states become a linear superposition of the bare states. Therefore, it is convenient to introduce symmetric and antisymmetric superpositions 
\begin{align}
\ket u &= \alpha\ket 3 +\beta\ket 1 ,\nonumber\\
\ket w &= \beta\ket 3 -\alpha\ket 1 ,\label{e16}
\end{align}
where
\begin{equation}
\alpha = \sqrt{\frac{\gamma_{1}}{\gamma_{1}+\gamma_{3}}} ,\quad \beta = \sqrt{\frac{\gamma_{3}}{\gamma_{1}+\gamma_{3}}} .\label{e17}
\end{equation}
Physically, the states $\ket u$ and $\ket w$ are the effective states between which the quantum dot evolves when interacting with the phonon bath.

It is then easily verified that in terms of the superposition states (\ref{e16}) the master equation (\ref{e13}) assumes the simplified form with
\begin{equation}
\tilde{H}_{0} = \Delta A_{22} +\Omega_{w}\left(A_{2w}+A_{w2}\right) +\Omega_{u}\left(A_{2u}+A_{u2}\right) ,\label{e18}
\end{equation}
where
\begin{equation}
\Omega_{w} = \beta\Omega_{3} -\alpha\Omega_{1} ,\quad
\Omega_{u} = \alpha\Omega_{3} +\beta\Omega_{1} ,\label{e19}
\end{equation}
the dissipative phonon-reservoir part ${\cal L}_{p}\rho$ reduced to
\begin{align}
{\cal L}_{p}\rho &= (\bar{n}+1)\gamma\left(2A_{w2}\rho A_{2w} -A_{22}\rho -\rho A_{22}\right)\nonumber\\
&+\bar{n}\gamma\left(2A_{2w}\rho A_{w2} -A_{ww}\rho -\rho A_{ww}\right) ,\label{e20}
\end{align}
in which $\gamma =\gamma_{1}+\gamma_{3}$.

The Liouvillians ${\cal L}_{s2}\rho$ and ${\cal L}_{s3}\rho$, representing dissipation due to the radiative spontaneous emission, written in basis of the superposition states $\ket w$ and $\ket u$ take the form
\begin{align}
{\cal L}_{s2}\rho &= \alpha^{2}\Gamma_2\left(\left[A_{w2}\rho, A_{2w}\right] +\left[A_{w2},\rho A_{2w}\right]\right)\nonumber\\
&+ \beta^{2}\Gamma_2\left(\left[A_{u2}\rho, A_{2u}\right] +\left[A_{u2},\rho A_{2u}\right]\right)\nonumber\\
&-\alpha\beta \Gamma_{2}\left(\left[A_{w2}\rho, A_{2u}\right] +\left[A_{w2},\rho A_{2u}\right] +{\rm H.c.}\right) ,\label{e21}
\end{align}
and
\begin{align}
{\cal L}_{s3}\rho &= \beta^{2}\Gamma_{3}\left(\left[A_{2w}\rho, A_{w2}\right] +\left[A_{2w},\rho A_{w2}\right]\right)\nonumber\\
&+ \alpha^{2}\Gamma_{3}\left(\left[A_{2u}\rho, A_{u2}\right] +\left[A_{2u},\rho A_{u2}\right]\right)\nonumber\\
&+\alpha\beta \Gamma_{3}\left(\left[A_{2w}\rho, A_{u2}\right] +\left[A_{2w},\rho A_{u2}\right] +{\rm H.c.}\right) .\label{e21n}
\end{align}

\begin{figure}[h]
\begin{center}
\begin{tabular}{c}
\includegraphics[width=0.65\columnwidth]{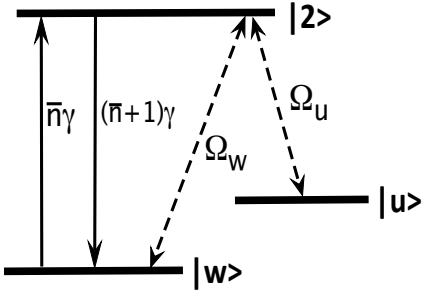}
\end{tabular}
\end{center}
\caption{The three-level quantum dot system in the superposition state basis $\{\ket 2,\ket w,\ket u\}$. In the superposition state basis the system is equivalent to a $\Lambda$-type system in which the phonon bath couples exclusively to the $\ket 2\leftrightarrow \ket w$ transition as an ordinary harmonic oscillator thermal bath. The transition $\ket 2\leftrightarrow \ket u$ is not affected by the phonon bath. The transition $\ket 2\leftrightarrow \ket u$ is driven by the laser with an effective Rabi frequency $\Omega_{u}$, whereas the transition $\ket 2\leftrightarrow \ket w$ is driven with an effective Rabi frequency $\Omega_{w}$.}
\label{fig3}
\end{figure}

It is seen from Eq.~(\ref{e20}) that the phonon reservoir couples only to the $\ket 2\leftrightarrow\ket w$ transition. The superposition state $\ket u$ is completely decoupled from the phonon bath. There is, however, a strong coherent coupling between $\ket u$ and $\ket 2$ with an effective Rabi frequency $\Omega_{u}$, as shown in Fig.~\ref{fig3}. The laser field couples the states $\ket w$ and $\ket 2$ with an effective Rabi frequency~$\Omega_{w}$.  

Figure~\ref{fig3} clearly illustrates that the three-level QD system interacting with the phonon bath effectively behaves as a $\Lambda$-type system. It is particularly well seen in the properties of the radiative damping. The Liouvillian ${\cal L}_{s2}\rho$ has the same form as the Liouvillian for a three-level $\Lambda$ type system with cross coupled transitions. It is easy to see, the first term in Eq.~(\ref{e21}) describes spontaneous decay from the level $\ket 2$ to level $\ket w$ with the rate $\alpha^{2}\Gamma_2$. The second term describes spontaneous decay from the level $\ket 2$ to level $\ket u$ with the rate $\beta^{2}\Gamma_2$. Clearly, the Liouvillian (\ref{e21}) describes dissipation of the three-level $\Lambda$ type system with the upper level $\ket 2$ and two ground levels $\ket w$ and $\ket u$. The third term in Eq.~(\ref{e21}) describes the dissipative cross coupling between these transitions with an amplitude $\alpha\beta\Gamma_{2}$.
What is more interesting that the spontaneous emission from the level $\ket 3$ has the same effect as an incoherent pumping of the $\Lambda$ system. It is seen from Eq.~(\ref{e21n}) that spontaneous emission from the level $\ket 3$ excites transitions from the ground levels $\ket w$ and $\ket u$ to the upper level $\ket 2$ with rates $\beta^{2}\Gamma_{3}$ and $\alpha^{2}\Gamma_{3}$, respectively. Thus, it represents a process of an incoherent pumping of the $\Lambda$ system. We may conclude, that the role of the radiative spontaneous emission in the ladder system interacting with a phonon bath is analogous to that of the dissipation and an incoherent driving of the $\Lambda$-type system with cross coupled transitions.
 
There are two interesting limits for the effective Rabi frequency $\Omega_{w}$: $\Omega_{w}=0$ and $\Omega_{w}=\Omega_{u}$. The limit $\Omega_{w}=0$ corresponds to $\Omega_{1}/\Omega_{3} =\beta/\alpha$, whereas the case $\Omega_{w}=\Omega_{u}$ corresponds to $\Omega_{1}/\Omega_{3} =(\beta -\alpha)/(\alpha +\beta)$.

\subsection{The case $\Omega_{w}=0$}

When $\Omega_{1}/\Omega_{3} =\beta/\alpha\equiv \sqrt{\gamma_{3}/\gamma_{1}}$, we see from Eq.~(\ref{e19}) that the effective Rabi frequency $\Omega_{w}$ is equal to zero. Under such condition, the laser field couples exclusively to the $\ket 2\leftrightarrow \ket u$ transition. 

The absence of the driving field on the $\ket w \leftrightarrow \ket 2$ transition makes the system to behave as a three-level $V$-type rather than the $\Lambda$-type system.
To demonstrate this analogy, we diagonalize the Hamiltonian $\tilde{H}_{0}=\Delta A_{22} +\Omega_{u}(A_{2u}+A_{u2})$ and find that the eigenstates of teh Hamiltonian, the so-called semiclassical dressed states are
\begin{align}
\ket m &= \sin\theta \ket 2 +\cos\theta \ket u ,\nonumber\\
\ket n &= \cos\theta \ket2 -\sin\theta \ket u ,\label{e23a}
\end{align}
where
\begin{equation}
\cos^{2}\theta =\frac{1}{2}-\frac{\Delta}{2\Omega} ,\label{e24f}
\end{equation}
and $\Omega=\sqrt{\Delta^{2}+4\Omega_{u}^{2}}$.

In the space spanned by the dressed states $\ket w$, $\ket m$ and $\ket n$, the master equation (\ref{e13}) takes the form
\begin{align}
\frac{\partial}{\partial t}{\rho} &= -i[\tilde{H}_{0},\rho] \nonumber\\
&+\sum_{k=m,n}\gamma_{k}\left(\left[A_{wk}\rho, A_{kw}\right] +\left[A_{wk},\rho A_{kw}\right]\right)\nonumber\\
&+\frac{\bar{n}}{\bar{n}+1}\sum_{k=m,n}\gamma_{k}\left(\left[A_{kw}\rho, A_{wk}\right] +\left[A_{kw},\rho A_{wk}\right]\right)\nonumber\\
&-\sqrt{\gamma_{m}\gamma_{n}}\left(\left[A_{wm}\rho, A_{nw}\right]\!+\!\left[A_{wm},\rho A_{nw}\right] +{\rm H.c.}\right) ,\label{e13u}
\end{align}
where the Hamiltonian $\tilde{H}_{0}$ can be written in the form
\begin{align}
\tilde{H}_{0} &= \frac{1}{2}(\Delta +\Omega)A_{mm} +\frac{1}{2}(\Delta -\Omega)A_{nn} ,
\end{align}
with 
\begin{align}
\gamma_{m}=2(\bar{n}+1)\gamma\sin^{2}\theta,\quad  \gamma_{n}=2(\bar{n}+1)\gamma\cos^{2}\theta .
\end{align}
The second term in Eq.~(\ref{e13u}) describes decay processes from the states $\ket m$ and $\ket n$ to the state $\ket w$ occurring with rates $\gamma_{m}$ and $\gamma_{n}$, respectively. The third term describes an incoherent pumping from state $\ket w$ to states $\ket m$ and $\ket n$, whereas the last term describes cross-correlations between the decay processes with rate $\sqrt{\gamma_{m}\gamma_{n}}$.
\begin{figure}[h]
\begin{center}
\begin{tabular}{c}
\includegraphics[width=0.7\columnwidth]{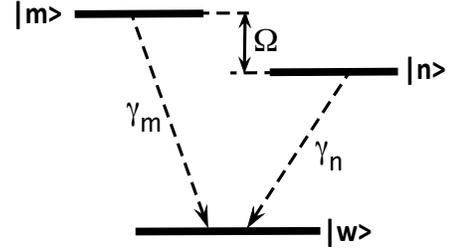}
\end{tabular}
\end{center}
\caption{Energy levels of the QD system in the dressed states basis $\{\ket w,\ket m,\ket n\}$. The system is equivalent to a non-degenerate $V$-type system with the upper dressed states $\ket m$ and $\ket n$ decaying to the lower state $\ket w$ with rates~$\gamma_{m}$ and $\gamma_{n}$, respectively. }
\label{fig2a}
\end{figure}

Figure~\ref{fig2a} shows the energy level structure of the QD system interacting with the phonon bath and driven by a laser field with the effective Rabi frequency $\Omega_{w}=0$. Clearly, the dynamics of the system are completely equivalent to those of an undriven three-level $V$-type system with non-degenerate transitions of the frequency difference $\Omega$, the upper states decaying with unequal rates $\gamma_{m}$ and $\gamma_{n}$, and the transitions cross-correlated with the rate~$\sqrt{\gamma_{m}\gamma_{n}}$.

\subsection{The case $\Omega_{w}=\Omega_{u}$}

When $\Omega_{w}$ and $\Omega_{u}$ are both diffrent from zero, the lower states $\ket w$ and $\ket u$ are resonately coupled to the upper state $\ket 2$ by the laser field with the effective Rabi frequencies $\Omega_{w}$ and $\Omega_{u}$, respectively. It is well known that in this configuration, the system is optically pumped into coherent superpositions of the two lower states, so-called bright and dark states~\cite{ao76,gw78,ea96,dk82,fs05}
\begin{align}
\ket b &= \left(\Omega_{w}\ket w +\Omega_{u}\ket u\right)/\sqrt{\Omega^{2}_{w}+\Omega^{2}_{u}} ,\nonumber\\
\ket d &= \left(\Omega_{u}\ket w -\Omega_{w}\ket u\right)/\sqrt{\Omega^{2}_{w}+\Omega^{2}_{u}} .\label{e37a}
\end{align}
In the case of $\Omega_{w}=\Omega_{u}$, the states (\ref{e37a}) simplify to 
\begin{align}
\ket b &=\frac{1}{\sqrt{2}}\left[\left(\alpha+\beta\right)\ket 3 -\left(\alpha-\beta\right)\ket 1\right] ,\nonumber\\
\ket d &=\frac{1}{\sqrt{2}}\left[\left(\alpha-\beta\right)\ket 3 +\left(\alpha+\beta\right)\ket 1\right] .\label{e37b}
\end{align}
Comparing Eq.~(\ref{e37b}) with Eq.~(\ref{e16}), we see that the states $\ket b, \ket d$ and $\ket u, \ket w$ are similar coherent superpositions of the bare states $\ket 1$ and $\ket 3$. However, their coherence properties are mutually exclusive in the sense that under a condition the states $\ket u$ and $\ket w$ are maximally coherent the states $\ket b$ and $\ket d$ are reduced to bare states, and vice-versa.  In particular, when $\alpha=\beta$ the states $\ket u,\ket w$ reduce to superposition states with maximal coherence between $\ket 1$ and $\ket 3$, whereas the states $\ket b, \ket d$ reduce to bare states $\ket 3, \ket 1$ with no coherence between them. Similarly, when $\alpha\gg\beta$ the superposition states $\ket u$ and $\ket w$ reduce to the bare states $\ket 3$ and $\ket 1$, respectively, whereas the states $\ket b$ and $\ket d$ reduce to superposition states with maximal coherence. 
  
Thus, in the case of $\Omega_{w}=\Omega_{u}$, the effect of the driving laser field is to destroy the coherent superpositions between the states $\ket 1$ and $\ket 3$ induced by the phonon bath.

\section{Population distribution}\label{sec3}

All the effects of the phonon reservoir discussed in the previous section should be reflected in the distribution of the populations between the energy levels of the QD. The usual way to study the population distribution is to derive equations of motion for the density matrix elements using the master equation of a given system.

Since we are interested in the population distribution and coherences between the superposition states, we determine the density matrix elements in the basis $\{\ket 2,\ket w,\ket u\}$. Within this basis, the resulting density matrix elements satisfy equations that show considerably less coupling. Including both, phonon-bath and radiative dampings it is straightforward to show that the density matrix elements obey the following set of coupled differential equations
\begin{align}
\dot{\rho}_{22} =& -2\!\left[\Gamma_{2}+(\bar{n}+1)\gamma\right]\rho_{22} +2\!\left(\beta^{2}\Gamma_{3}+\bar{n}\gamma\right)\rho_{ww} \nonumber\\
&+2\alpha^{2}\Gamma_{3}\rho_{uu} +2\alpha\beta\Gamma_{3}(\rho_{wu}+\rho_{uw}) \nonumber\\
& +i\Omega_{u}\left(\rho_{2u}-\rho_{u2}\right) +i\Omega_{w}\left(\rho_{2w}-\rho_{w2}\right) ,\nonumber\\
\dot{\rho}_{ww} =& -2\!\left(\beta^{2}\Gamma_{3}+\bar{n}\gamma\right)\rho_{ww} +2\!\left[\alpha^{2}\Gamma_{2}+(\bar{n}+1)\gamma\right]\rho_{22} \nonumber\\
& -\alpha\beta\Gamma_{3}(\rho_{wu}+\rho_{uw}) -i\Omega_{w}\left(\rho_{2w}-\rho_{w2}\right) ,\nonumber\\
\dot{\rho}_{uu} =& -2\alpha^{2}\Gamma_{3}\rho_{uu}+2\beta^{2}\Gamma_{2}\rho_{22} -\alpha\beta\Gamma_{3}(\rho_{wu}+\rho_{uw}) \nonumber\\
&-i\Omega_{u}\left(\rho_{2u} -\rho_{u2}\right) ,\nonumber\\
\dot{\rho}_{2w} =& -\left[\Gamma_{2}+\beta^{2}\Gamma_{3}+\left(2\bar{n}+1\right)\gamma +i\Delta\right]\rho_{2w}  \nonumber\\
&-\alpha\beta\Gamma_{3}\rho_{2u} -i\Omega_{u}\rho_{uw} +i\Omega_{w}\left(\rho_{22}-\rho_{ww}\right) , \nonumber\\
\dot{\rho}_{2u} =& -\left[\Gamma_{2}+\alpha^{2}\Gamma_{3}+\left(\bar{n}+1\right)\gamma +i\Delta\right]\rho_{2u} \nonumber\\
&-\alpha\beta\Gamma_{3}\rho_{2w} -i\Omega_{w}\rho_{wu} +i\Omega_{u}\left(\rho_{22}-\rho_{uu}\right) , \nonumber\\
\dot{\rho}_{wu} =& -\alpha\beta\Gamma_{3}-\left(\Gamma_{3}+\bar{n}\gamma\right)\rho_{wu} -\alpha\beta\left(2\Gamma_{2}-\Gamma_{3}\right)\rho_{22} \nonumber\\
& +i\Omega_{u}\rho_{w2} -i\Omega_{w}\rho_{2u} .\label{e22}
\end{align}
Equations (\ref{e22}) plus the equations for $\rho_{w2},\rho_{u2},\rho_{uw}$ form a set of nine coupled differential equations with constant coefficients. These equations are cumbersome because of a complicated coupling, which exists in general between all three pairs of the states, $\ket w\leftrightarrow \ket 2$, $\ket 2\leftrightarrow \ket u$ and $\ket u\leftrightarrow \ket w$. The equations can be simplified substantially for particular choices of the parameters, such as $\Gamma_{2},\Gamma_{3}\approx 0$. In physical terms, it would correspond to the situation where the radiation modes occupy only a small fraction of modes surrounding the QD.

Suppose first that $\Gamma_{2},\Gamma_{3}\approx 0$ that the system is affected solely by the phonon bath. It is seen from Eq.~(\ref{e22}) that in this case the population $\rho_{uu}$ does not decay. In other words, the state $\ket u$ appears as a decoherence free state. The decay occurs only from the state $\ket 2$ to the state $\ket w$. Thus, owing the equivalence of the system with a non-degenerate three-level $V$-type system, we may conclude that in the space spanned by the states $\ket 2$, $\ket w$ and $\ket u$, the state $\ket 2$ coincides with a symmetric state which decays to the ground state $\ket w$, whereas the state $\ket u$ coincides with a non-decaying asymmetric state coherently coupled to the state $\ket 2$ with the amplitude~$\Omega_{u}$.

Further simplification can be made by choosing the Rabi frequencies such as $\Omega_{w}= 0$ or~$\Omega_{w}=\Omega_{u}$. In the first case of $\Omega_{w}=0$, the set of the differential equations (\ref{e22}) splits into two independent sets, one involving four equations
\begin{align}
\dot{\rho}_{22} =& -2(\bar{n}+1)\gamma\rho_{22} +2\bar{n}\gamma\rho_{ww} +i\Omega_{u}\left(\rho_{2u}-\rho_{u2}\right) ,\nonumber\\
\dot{\rho}_{ww} =& -2\bar{n}\gamma\rho_{ww} +2(\bar{n}+1)\gamma\rho_{22} ,\nonumber\\
\dot{\rho}_{uu} =& -i\Omega_{u}\left(\rho_{2u} -\rho_{u2}\right) ,\nonumber\\
\dot{\rho}_{2u} =& -\left[\left(\bar{n}+1\right)\gamma +i\Delta\right]\rho_{2u} +i\Omega_{u}\left(\rho_{22}-\rho_{uu}\right) ,\label{e22a}
\end{align}
and the other involving two coupled equations
\begin{align}
\dot{\rho}_{2w} =& -\left[\left(2\bar{n}+1\right)\gamma +i\Delta\right]\rho_{2w} -i\Omega_{u}\rho_{uw} ,\nonumber\\
\dot{\rho}_{wu} =& -\bar{n}\gamma\rho_{wu} +i\Omega_{u}\rho_{w2} .\label{e22b}
\end{align}
Equations (\ref{e22a}) and (\ref{e22b}) can be solved strictly. The system of equations (\ref{e22a}) can be transformed using the Laplace transform method into a set of algebraic equations which can be solve, for example, by matrix inversion. We will consider the time evolution of the density matrix elements in Sec.~\ref{sec5}. Here, we focus on the steady-state solutions.
The set (\ref{e22a}) can have a nonzero steady-state solution, while the steady-state solution of Eq.~(\ref{e22b}) is zero.

The steady-state can be a pure state when the phonon bath is at zero temperature $(\bar{n}=0)$ and the Rabi frequency $\Omega_{w}$ is made zero by a suitable choice of the ratio $\Omega_{3}/\Omega_{1}=\gamma_{1}/\gamma_{3}$. In the case when $\bar{n}=0$ and $\Omega_{w}=0$, we easily find that $\rho_{ww}=1$. We should point out the fact that the decay of the system to the pure state does not depend on the particular values of the decay rates and is valid for $\gamma_{1}=\gamma_{3}$ as well as for $\gamma_{1}\neq \gamma_{3}$.

In the case of $\Omega_{w}=0$, the steady-state solution of Eq.~(\ref{e22}) is given by
\begin{equation}
\rho_{ww}=\frac{\bar{n}+1}{3\bar{n}+1} ,\quad \rho_{uu}=\rho_{22}=\frac{\bar{n}}{3\bar{n}+1} ,
\end{equation}
from which we see that the populations are independent of $\gamma_{1}$ and $\gamma_{3}$. Moreover, the populations are independent of~$\Omega_{u}$. Note that there is no population inversion between $\ket 2$ and~$\ket w$. Thus, there is no population inversion in the basis of the superposition states. However, there is population inversion between the bare states. It is straightforward to show that the steady-state populations of the bare states are
\begin{equation}
\rho_{33}=\frac{\bar{n}+\beta^{2}}{3\bar{n}+1} ,\quad \rho_{22}=\frac{\bar{n}}{3\bar{n}+1} ,\quad \rho_{11}=\frac{\bar{n}+\alpha^{2}}{3\bar{n}+1} .\label{e24}
\end{equation}
It is apparent from Eq.~(\ref{e24}) that always $\rho_{33}>\rho_{22}$. Hence there is always a population inversion between the levels $\ket 3$ and $\ket 2$. The maximum inversion is reached at $\bar{n}=0$ and decreases with an increasing $\bar{n}$. The steady-state also posses a large two-photon coherence
\begin{equation}
\rho_{13} =-\frac{\alpha\beta}{3\bar{n}+1} .\label{e25}
\end{equation}
Thus, the system decays to a strongly correlated state in which the population distribution no longer obeys a Boltzmann distribution. Moreover, for $\bar{n}=0$ the correlated state is a pure state. It is particularly seen when one calculate the purity factor
\begin{equation}
{\rm Tr}(\rho^{2}) = \frac{3\bar{n}^{2}+2\bar{n}+1}{\left(3\bar{n}+1\right)^{2}} ,
\end{equation}
from which it is evident that ${\rm Tr}(\rho^{2}) =1$ for $\bar{n}=0$.

Furthermore, if $\beta>\alpha$ then $\rho_{33}>\rho_{11}$ for any $\bar{n}$, corresponding to a population inversion between levels $\ket 3$ and $\ket 1$. Thus, in the case of $\gamma_{3}>\gamma_{1}$, we can have not only an one-photon inversion between levels $\ket 3$ and $\ket 2$ but also a two-photon inversion between levels $\ket 3$ and~$\ket 1$.
\begin{figure}[h]
\begin{center}
\begin{tabular}{c}
\includegraphics[width=\columnwidth]{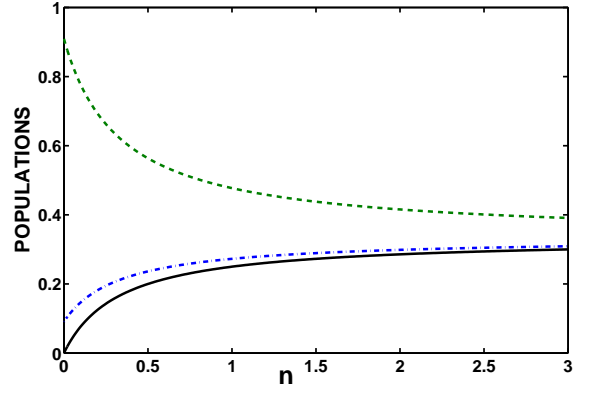}
\end{tabular}
\end{center}
\caption{(Color online) Steady-state populations of the bare energy levels of the quantum dot: $\rho_{22}$ (black solid line), $\rho_{33}$ (green dashed line), $\rho_{11}$ (blue dashed-dotted line) for $\gamma_{1}=\gamma_{0}$, $\gamma_{3}=10\gamma_{0}$, $\Omega_{3}/\Omega_{1}=\sqrt{\gamma_{1}/\gamma_{3}}$, $(\Omega_{w}=0)$, and $\Delta=5\gamma_{0}$, where $\gamma_{0}$ is a mean damping rate and we take $\gamma_{0}=1$ throughout.}
\label{fig3a}
\end{figure}

The above considerations are illustrated in Fig.~\ref{fig3a}, which shows the populations of the bare states as a function of $\bar{n}$ for the case of $\Omega_{w}=0$. The one-photon $(\rho_{33}>\rho_{22})$ and two-photon $(\rho_{33}>\rho_{11})$ inversions are clearly seen to occur for all~$\bar{n}$. For $\bar{n}=0$ the inversions are large but fall as $\bar{n}$ increases.
\begin{figure}[h]
\begin{center}
\begin{tabular}{c}
\includegraphics[width=0.9\columnwidth]{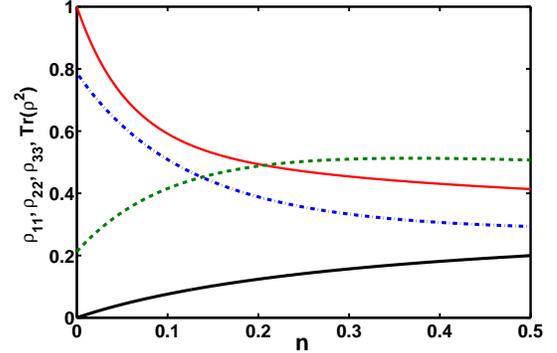}
\end{tabular}
\end{center}
\caption{(Color online) Steady-state populations of the bare energy levels of the quantum dot; $\rho_{22}$ (black solid line), $\rho_{33}$ (green dashed line), $\rho_{11}$ (blue dashed-dotted line) for the case of $\Omega_{w}=\Omega_{u}$ with $\gamma_{1}=\gamma_{0}$, $\gamma_{3}=10\gamma_{0}$, $\Omega_{3}=5\gamma_{0}$, and $\Delta=5\gamma_{0}$. The upper solid red line shows ${\rm Tr}(\rho^{2})$.}
\label{fig4}
\end{figure}

Figure~\ref{fig4} show the variation of the populations with $\bar{n}$ for the situation when $\Omega_{w}=\Omega_{u}$. In that case simple analytical results for the density matrix elements are no longer possible. The populations are obtained by solving numerically the set of coupled differential equations (\ref{e22}) in the steady-state limit, $(\dot{\rho}_{ij}=0)$. Once again we notice that in the limit of $\bar{n}=0$ the system is in a pure state showing that coherences are present when $\Omega_{w}=\Omega_{u}$. It is interesting that the pure state does not coincide with the state $\ket w$ to which the system decays when $\Omega_{w}=0$. 

It is straightforward to show using Eqs.~(\ref{e37b}) and (\ref{e22}) that the population of the dark state $\ket d$ satisfies the equation of motion
\begin{equation}
\dot{\rho}_{dd} = -\bar{n}\gamma\rho_{dd} +(\bar{n}+1)\rho_{22} -\bar{n}\gamma\left(\rho_{bd} +\rho_{db}\right) .
\end{equation}
It is apparent that the evolution of the population $\rho_{dd}$ is independent of the Rabi frequency and in the limit of $\bar{n}=0$ the population increases with the rate $\gamma$ due to the transfer of the population from the state $\ket 2$. Therefore, in the steady-state~$\rho_{dd}=1$, which means that the system decays to the steady-state which is the pure state $\ket d$.

One can notice from Fig.~\ref{fig4} that in contrast to the case $\Omega_{w}=0$ there is no two-photon population inversion at $\bar{n}=0$, i.e., $\rho_{11}>\rho_{33}$. However, the populations can be inverted for a finite $\bar{n}$. It is readily understood if we refer to the state $\ket d$, Eq.~(\ref{e37b}), from which it is clear that the amplitude $(\alpha-\beta)$ of the state~$\ket 3$ is always smaller than the amplitude $(\alpha+\beta)$ of the state $\ket 1$. Evidently, $\rho_{11}$ is always larger than $\rho_{33}$ independent of the ratio $\gamma_{3}/\gamma_{1}$ and the Rabi frequencies of the driving field.

It is worth noting that the two cases of $\Omega_{w}=0$ and $\Omega_{w}=\Omega_{u}$ are not directly comparable, although the pure states, respectively $\ket w$ and $\ket d$ to which the system evolves at $\bar{n}=0$ are similar superpositions of the states $\ket 1$ and $\ket 3$. This conclusion is especially evident if we examine the case of $\gamma_{1}=\gamma_{3}$, for which $\alpha=\beta=1/\sqrt{2}$. Under this condition, we find from Eqs.~(\ref{e16}) and (\ref{e37b}) that the state $\ket w$ reduces to a superposition state with maximal coherence $(|\rho_{13}|=1/2)$, whereas the state $\ket d$ reduces to the bare state $\ket 1$.

When $\bar{n}>0$, the populations $\rho_{11}$ and $\rho_{33}$ become inverted. This shows that for $\bar{n}>0$ the competing effects of the phonon bath dominates over the laser field and cause a switch of the evolution of the system from that between the states $\ket b, \ket d$ and $\ket 2$ to that occurring between the states $\ket w, \ket u$ and $\ket 2$. In other words, for $\bar{n}>0$, the population distribution switches to that of the states $\ket w$ and $\ket u$ mediated by the phonon bath. It follows that there is an evident competition between the phonon bath and the driving field. The phonon bath leads to strong coherence between the bare states with no coherences between the states $\ket w$ and $\ket u$. Inversely, the driving fields leads to strong coherences between the superposition states~$\ket w$ and $\ket u$ which results in a reduced coherence between the bare state basis.

\section{Squeezed states and photon correlations}\label{sec4}

Since the steady-state of the system can posse a large coherence $\rho_{13}$, the radiation field emitted by the quantum dot may experience significant squeezing under the influence of the phonon bath. In order to consider the ability of the system to generate squeezed light, we calculate the quantities that characterize the squeezing of the emitted light. Squeezed states of light are associated with the requirement that the normally ordered variance of the electric-field quadrature component $E_{\phi}$ is negative. The normally ordered variance of the quadrature component $E_{\phi}$ is defined by~\cite{gv12,or11,gv13}
\begin{align}
\langle :(\Delta E_{\phi})^{2}:\rangle &= \langle(\Delta E^{(+)})^{2}\rangle {\rm e}^{2i\phi}
+\langle(\Delta E^{(-)})^{2}\rangle {\rm e}^{-2i\phi}\nonumber\\
& +2\langle\Delta E^{(-)}E^{(+)}\rangle ,
\end{align}
where $\Delta E^{(\pm)} =E^{(\pm)} -\langle E^{(\pm)}\rangle$ and $E^{(+)}\, (E^{(-)})$ is the positive (negative) frequency component of the electromagnetic field. The normally ordered variance is directly measurable in schemes involving homodyne or heterodyne detection and gives information about relative squeezing of the field at particular phase $\phi$. The phase $\phi$ is the phase difference between the driving field, or the initial dipole moment of the system, and the local oscillator in the detector.

The positive frequency component of the radiated field from a three-level cascade system and detected at a point $\vec{r}$ in the far field zone can be expressed in terms of the dipole operators as
\begin{equation}
E^{(+)}(\vec{r},t) = \psi(\vec{r}\,)\left(A_{23}(t) +A_{21}(t)\right) ,
\end{equation}
where $\psi(\vec{r}\,)$ is a geometrical factor which depends on magnitudes of the transition dipole moments and their polarization in respect to the direction of observation $\vec{r}$. Therefore, the normally ordered variance can be expressed in terms of the density matrix elements as
\begin{align}
\langle :\!(\Delta E_{\phi})^{2}\!:\rangle &= \psi^{2}(\vec{r}\,)\!\left\{\rho_{22}\!+\!\rho_{33} -\frac{1}{2}\!\left[(\rho_{32}+\rho_{21}){\rm e}^{i\phi}\!+\!{\rm c.c.}\right]\right. \nonumber\\
&+\left. \frac{1}{2}\left(\rho_{31}{\rm e}^{2i\phi} +\rho_{13}{\rm e}^{-2i\phi}\right)\right\} .
\end{align}

Since in the steady-state $\rho_{32}=\rho_{21}=0$, the normally ordered variance is governed by the difference between the sum of the excited levels population and the real part of the two-photon coherence. Therefore, $\rho_{13}\neq 0$ is necessary to produce squeezing. We note that the requirement that $\rho_{13}\neq 0$ is necessary but not sufficient for squeezing. Using Eqs.~(\ref{e24}) and (\ref{e25}), we find that in the steady-state the normally ordered variance takes the form
\begin{equation}
\langle :(\Delta E_{\phi})^{2}:\rangle = \psi^{2}(\vec{r}\,)\frac{2\bar{n}+\beta\left[\beta -\alpha\cos(2\phi)\right]}{3\bar{n}+1} .\label{e29}
\end{equation}
Clearly this variance is always positive if~$\alpha<\beta$. If $\alpha>\beta$, the variance can be negative indicating that the system radiates squeezed light. Hence $\alpha>\beta$ is the general condition for squeezing in the system that squeezing is possible only if there is no population inversion between the levels $\ket 1$ and $\ket 3$.
The minimum value of the variance corresponding to maximum squeezing reached in the quadrature component $E_{0}\, (\phi=0)$ when $\bar{n}=0$, in which case the state of the system is a pure state. However, it is interesting to note that the optimum squeezing occurs not at the largest value of the coherence $\rho_{13}=-1/2$. It is easily verified from Eq.~(\ref{e29}) that in the case of $\rho_{13}=-1/2$, corresponding to $\alpha=\beta =1/\sqrt{2}$, the variance $\langle :(\Delta E_{0})^{2}:\rangle =0$. Normally, we would expect squeezing to attain its maximum value when the coherence is maximal. The reason is that the population $\rho_{33}$ also depends on $\alpha$ and attains a minimum value not for $\alpha=1/\sqrt{2}$ but for $\alpha=1$.

A careful examination of Eq.~(\ref{e29}) reveals that the optimum squeezing occurs for $\bar{n}=0$ and $\beta/\alpha=\sqrt{2}-1$, when it reaches the value
\begin{equation}
\langle :(\Delta E_{0})^{2}:\rangle/\psi^{2}(\vec{r}\,) = -\frac{1}{2}\left(\sqrt{2}-1\right) .
\end{equation}
This is the maximum amount of squeezing possible in a three-level system in the cascade configuration~\cite{df94}. For $\bar{n}\neq 0$, corresponding to a mixed state, the amount of squeezing is necessarily smaller. It follows from Eq.~(\ref{e16}) that the state corresponding to the optimum squeezing is of the form~\cite{df94}
\begin{equation}
\ket w = \sqrt{\frac{1}{2}\left(1-\frac{1}{\sqrt{2}}\right)}\ket 3 -\sqrt{\frac{1}{2}\left(1+\frac{1}{\sqrt{2}}\right)}\ket 1 .\label{e31}
\end{equation}
The state (\ref{e31}) corresponds to the two-photon coherence that is $\sqrt{2}$ smaller than its maximal value, $\rho_{13}=-1/(2\sqrt{2})$.
\begin{figure}[h]
\begin{center}
\begin{tabular}{c}
\includegraphics[width=\columnwidth]{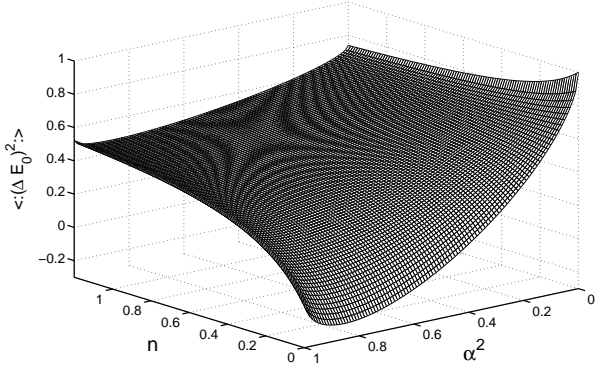}
\end{tabular}
\end{center}
\caption{Variation of the steady-state variance $\langle :(\Delta E_{0})^{2}:\rangle/\psi^{2}(\vec{r}\,)$ with $\bar{n}$ and $\alpha^{2}$ for $\Omega_{w}=0$.}
\label{fig5}
\end{figure}

The above considerations are illustrated in Fig.~\ref{fig5}, which shows the variation of $\langle :(\Delta E_{0})^{2}:\rangle$ with $\bar{n}$ and $\alpha^{2}$. Squeezing is seen to occur in a restricted range of the parameters, it is confined to small~$\bar{n}$ and $\alpha^{2}>1/2$.

\begin{figure}[h]
\begin{center}
\begin{tabular}{c}
\includegraphics[width=\columnwidth]{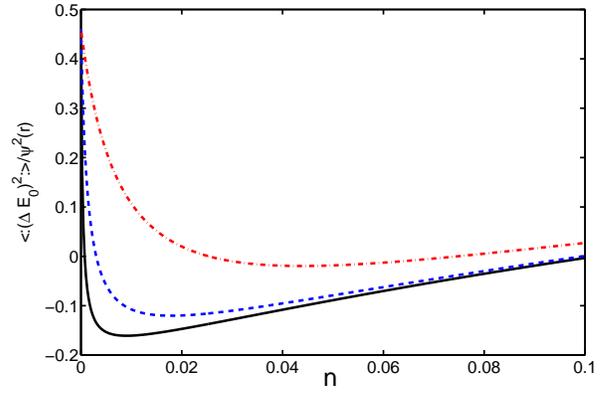}
\end{tabular}
\end{center}
\caption{(Color online) The variance $\langle :(\Delta E_{0})^{2}:\rangle/\psi^{2}(\vec{r}\,)$ as a function of $\bar{n}$ for the case $\Omega_{w}=\Omega_{u}$ with $\gamma_{1}=5\gamma_{0}, \gamma_{3}=\gamma_{0}$\, $(\alpha=0.9129)$ and different Rabi frequencies $\Omega_{3}$: $\Omega_{3}=0.1\gamma_{0}$ (solid black line), $\Omega_{3}=0.2\gamma_{0}$ (dashed blue line), $\Omega_{3}=0.5\gamma_{0}$ (dashed-dotted red line).}
\label{fig6}
\end{figure}

Figure~\ref{fig6} shows the variation of the steady-state variance $\langle :(\Delta E_{0})^{2}:\rangle/\psi^{2}(\vec{r}\,)$ with $\bar{n}$ for the case of the symmetric driving with $\Omega_{w}=\Omega_{u}$. It is seen that the variance is positive at $\bar{n}=0$ so there is no squeezing. The reason for this is that the driving field coupled to both $\ket w \rightarrow \ket 2$ and $\ket u\rightarrow \ket 2$ transitions creates a coherence between the states $\ket w$ and $\ket u$ which diminishes or even can completely destroy the two-photon coherence~$\rho_{13}$ responsible for squeezing. However, as we have discussed in Sec.~\ref{sec3}, a finite temperature phonon bath $(\bar{n}\neq 0)$ can enhance the two-photon coherence and squeezing is recovered with the variance negative for small Rabi frequencies. It is as if the driving field is weak, the coherence is slaved to return to that between the states $\ket 1$ and $\ket 3$ extremely rapidly.  The effect of increasing the Rabi frequency is clearly to decrease the amount of squeezing and also to restrict further the range of $\bar{n}$ where it occurs.

In closing this section we would like to point out that the decay of the system to the two-photon correlated state results in a strong second-order coherence~\cite{mw95}. The coherence is determined by the normalized second-order correlation function~$g^{(2)}$, which for a three-level system in a ladder configuration is given by
\begin{equation}
g^{(2)} = \frac{\rho_{33}}{\left(\rho_{33}+\rho_{22}\right)^{2}} .
\end{equation}
For the case of $\Omega_{w}=0$, in which the system may decay to the pure two-photon state $\ket w$, we find using Eq.~(\ref{e24}) that
\begin{equation}
g^{(2)} = \frac{(3\bar{n}+1)(\bar{n}+\beta^{2})}{(2\bar{n}+\beta^{2})^{2}} .
\end{equation}

Certain interesting features of the second-order coherence function follow from this expression. First, we note that in the limit of~$\bar{n}=0$, in which the system decays to the pure state $\ket w$, the second-order correlation function reduces to $g^{(2)}=1/\beta^{2}$. Since $\beta\leq 1$, the stationary state always exhibits bunching effect, $g^{(2)}>1$. Moreover, for $\beta\ll 1$ the correlation function can reach very large values, clearly demonstrating the possibility for super-bunching~\cite{sz00}. Second, we note that $g^{(2)}$ decreases with increasing $\bar{n}$, which indicates that the stationary state with a large $\bar{n}$ is more "coherent" than that with a small~$\bar{n}$. Moreover, for $\bar{n}\neq 0$ a state of the system, which is a mixed state, may exhibits the nonclassical phenomenon of antibunching $(g^{(2)}<1)$ which can persist even in the limit of $\bar{n}\rightarrow\infty$, where $g^{(2)}=3/4$. Finally, we note the connection of squeezing with the phenomenon of bunching rather than with antibunching. From the general condition for squeezing, $\alpha^{2}>\beta^{2}$, or equivalently $\beta^{2}<1/2$, we find that squeezing corresponds to $g^{(2)}>2$. It follows that squeezing is connected to correlations stronger than that for a thermal state, $g^{(2)}=2$. The optimum squeezing for the pure state $\ket w$ was shown earlier to occurs for $\beta^{2}= (\sqrt{2}-1)/(2\sqrt{2})$, in which case $g^{(2)}\approx 7$. It is interesting and perhaps surprising that the stationary state with maximal coherence $\rho_{13}=-1/2$ is a thermal state. It occurs for $\alpha=\beta=1/\sqrt{2}$, for which~$g^{(2)}=2$.

\section{Quantum beats and population trapping}\label{sec5}

A useful way of illustrating the presence of coherence induced by the phonon bath is to consider the transient properties of the system. To do that, we calculate the time evolution of the total intensity of the radiation field spontaneously emitted by the system. For a multi-level system the intensity is sensitive to correlations between different transitions of the system and therefore it may be employed to manifest the existence of the correlations induced by the phonon bath between the two transitions of the ladder system. The time evolution of the radiation intensity, in particular, its oscillatory behaviour is a sensitive function of the initial conditions. In order to study this dependence we first consider the temporal behaviour of the populations of the energy states of the system. We then extend the discussion to the total radiation intensity of the emitted field. In the calculation of the radiation intensity we assume that the dynamics of the QD are affected solely by the phonon bath but the coupling between the QD and the electromagnetic vacuum is very weak as compared with the coupling between the QD and the phonon bath. 

The total radiation intensity is proportional to the total rate of energy loss from the excited states of the system, which in the case of a ladder three-level system, with~$\omega_{32},\omega_{21}\gg \Delta$, can be written~as
\begin{align}
&I(t) = \langle E^{(-)}(t)E^{(+)}(t)\rangle \nonumber\\
&\propto\!\langle(\!\sqrt{\Gamma_{3}}A_{32}(t)\!+\!\sqrt{\Gamma_{2}}A_{21}(t))(\!\sqrt{\Gamma_{3}}A_{23}(t)\!+\!\sqrt{\Gamma_{2}}A_{12}(t))\rangle \nonumber\\
&= \Gamma_{3}\rho_{33}(t) +\Gamma_{2}\rho_{22}(t) ,
\end{align}
where $\Gamma_{2}$ and $\Gamma_{3}$ are the damping rates at which the radiation is emitted from the states $\ket 3$ and $\ket 2$ to modes different than the phonon modes. We assume that $\Gamma_{2}=\Gamma_{3}\equiv \Gamma$ and $\Gamma$ is much smaller than the phonon damping rate $\gamma$, $(\Gamma\ll \gamma)$. 
It is seen that the intensity is proportional to the populations of the excited levels of the system. There is no contribution of the cross (interference) terms to the intensity. This property corresponds to properties of two transitions contributing independently to the intensity. Thus, in the case of the pure radiative damping of the ladder system the intensity of the radiated field tells us nothing about the correlations between the transitions.

However, the situation differs when in addition to the electromagnetic field the system is coupled to a dissipative phonon bath. In this case, quantum interference may occur due to the cross coupling between the two transitions created by the phonon bath. The correlations between the transitions are displayed in the transient properties of the populations.

In order to show this, we consider the time evolution of the populations for the case~$\Omega_{w}=0$. With the further assumption of a strong driving field, that is with $\Omega_{u}\gg \Delta,\bar{n}\gamma$, we find that to the first order in $\gamma/\Omega_{u}$ the temporal behaviour of the populations is
\begin{align}
\rho_{ww}(t) &= \frac{\bar{n}+1}{3\bar{n}+1} -\left[\frac{\bar{n}+1}{3\bar{n}+1}-\rho_{ww}(0)\right]{\rm e}^{-(3\bar{n}+1)\gamma t} \nonumber\\
&+\frac{(\bar{n}+1)\gamma}{2\Omega_{u}}\left[\rho_{22}(0) -\rho_{uu}(0)\right]{\rm e}^{-(\bar{n}+1)\gamma t}\sin(2\Omega_{u}t) ,\nonumber\\
\rho_{uu}(t) &= \frac{\bar{n}}{3\bar{n}+1} +\frac{1}{2}\left[\frac{\bar{n}+1}{3\bar{n}+1}-\rho_{ww}(0)\right]{\rm e}^{-(3\bar{n}+1)\gamma t} \nonumber\\
&-\frac{1}{2}\left[\rho_{22}(0)-\rho_{uu}(0)\right]{\rm e}^{-(\bar{n}+1)\gamma t}\cos(2\Omega_{u}t) \nonumber\\
&-\frac{\bar{n}\gamma}{2\Omega_{u}}\left[\rho_{ww}(0)-2\rho_{uu}(0)\right]{\rm e}^{-(\bar{n}+1)\gamma t}\sin(2\Omega_{u}t) ,\label{e47}
\end{align}
and the temporal behaviour of the coherence $\rho_{2u}(t)$ is given~by
\begin{align}
\rho_{2u}(t) &= i\left[\rho_{22}(0) -\rho_{uu}(0)\right]{\rm e}^{-(\bar{n}+1)\gamma t}\sin(2\Omega_{u}t) \nonumber\\
&-i\frac{\gamma}{\Omega_{u}}\left[(\bar{n}+1)-(3\bar{n}+1)\rho_{ww}(0)\right] \nonumber\\
&\times [{\rm e}^{-(3\bar{n}+1)\gamma t} -{\rm e}^{-(\bar{n}+1)\gamma t}\cos(2\Omega_{u}t)] .\label{e48}
\end{align}
We see that the time evolution of the populations and the coherence generally involves oscillatory and non-oscillatory terms. Depending on the choice of the initial state, quantum beats can be seen in the evolution of the populations. These quantum beats then can be seen in the radiation intensity, which in terms of $\rho_{ww}(t)$ and $\rho_{uu}(t)$ can be written~as
\begin{equation}
I(t) = \Gamma\!\left[\rho_{22}(t)\!+\!\alpha^{2}\rho_{uu}(t)\!+\!\beta^{2}\rho_{ww}(t)\!+\!2\alpha\beta {\rm Re}\rho_{uw}(t)\right] .\label{e49}
\end{equation}

We now examine the temporal behaviour of the populations for different initial conditions. First, we note that a non-zero coherence $\rho_{2u}$ can create quantum beats in the time evolution of the population of the state $\ket w$ even though the state is coupled to the state $\ket 2$ by the incoherent process. If $\rho_{ww}(0)=1$, the evolution of the populations is either exponential or oscillating, depending upon whether $\bar{n}=0$ or $\bar{n}\neq 0$. For $\bar{n}=0$ the temporal behaviour of the populations contains no oscillations. Only a nonzero number of phonons $(\bar{n}\neq 0)$ can lead to oscillations, quantum beats in the evolution of the populations. If $\bar{n}\neq 0$, the coherence $\rho_{2u}(t)$ is different from zero for any initial conditions. Thus, we expect the presence of quantum beats in the radiation intensity for any initial conditions if $\bar{n}\neq 0$. It is easily verified from Eqs.~(\ref{e47}) and (\ref{e48}) that for the initial condition $\rho_{ww}(0)=1$ and $\bar{n}=0$, the coherence $\rho_{2u}(t)=0$ for all times and the populations decay exponentially in time without any oscillations. Next, we note that the population $\rho_{uu}(t)$, but not $\rho_{ww}(t)$, contains an oscillatory term whose amplitude can be especially pronounced when $\rho_{22}(0)\neq \rho_{uu}(0)$. Otherwise, it shows oscillatory behaviour only when~$\bar{n}\neq 0$ and of a relatively small amplitude.

We now give illustrative examples of temporal behaves of the populations and the radiation intensity for a number of initial conditions.
\begin{figure}[h]
\begin{center}
\begin{tabular}{c}
\includegraphics[width=\columnwidth]{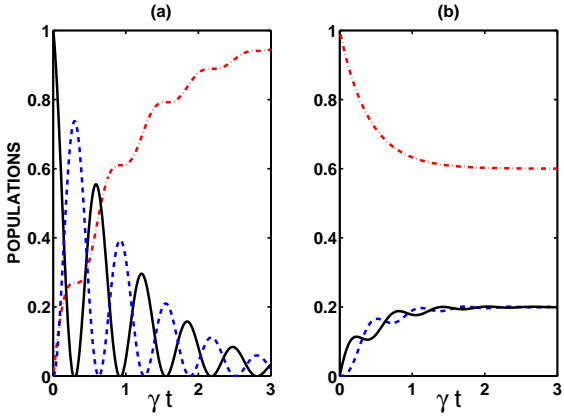}
\end{tabular}
\end{center}
\caption{(Color online) The transient behaviour of the populations $\rho_{22}(t)$ (black solid line), $\rho_{uu}(t)$ (blue dashed line) and $\rho_{ww}(t)$ (red dashed-dotted line) for $\Omega_{w}=0$, $\Omega_{u}=5\gamma_{0}$ and two sets of initial conditions: (a) $\rho_{22}(0)=1$, $\bar{n}=0$, (b) $\rho_{ww}(0)=1, \bar{n}=0.5$.}
\label{fig7}
\end{figure}

Figure~\ref{fig7} shows the time evolution of the populations for two different initial conditions and different $\bar{n}$. For the initial condition $\rho_{22}(0)=1$, illustrated in Fig.~\ref{fig7}(a), all the populations evolve in an oscillatory way. The pronounced sinusoidal oscillations (quantum beats) are clearly visible in the time evolution of the populations $\rho_{22}(t)$ and $\rho_{uu}(t)$, while the population~$\rho_{ww}(t)$ increases almost steady with oscillations at a relatively small amplitude. This oscillatory feature is associated with the fact that for the initial condition $\rho_{22}(0)=1$ the coherence $\rho_{2u}(t)$ is different from zero and oscillates at a large amplitude. A complete periodic depopulation of the states $\ket 2$ and $\ket u$ occurs, respectively, at times $t_{2}=(n+1)\pi/(2\Omega_{u})$ and $t_{u}=n\pi/(2\Omega_{u}),\, n=0,2,4,\ldots$. The periodic exchange of the population between the states $\ket 2$ and $\ket u$ is due to the coherent coupling attributable to the driving laser field whereas the steady increase the population $\rho_{ww}(t)$ is due to the decay process attributable to the interaction of the QD with the phonon bath. This periodic exchange of the population between the states $\ket 2$ and $\ket u$ continues until the QD decays to the superposition state $\ket w$.
Note an interesting feature that the maxima and minima of the population $\rho_{ww}(t)$ occur at times when the curves for the populations $\rho_{22}(t)$ and $\rho_{uu}(t)$ intersect, i.e. when $\rho_{22}(t)=\rho_{uu}(t)$. This clearly shows that the oscillations in $\rho_{ww}(t)$ are due to the coherence brought about by spontaneous (incoherent) transitions between the states $\ket w$ and $\ket 2$ induced by the phonon bath.

The situation differs when the system is initially at $t=0$ prepared in the state~$\ket w$. This is illustrated in Fig.~\ref{fig7}(b). The most obvious difference is that now oscillations are seen in the time evolution of only the populations $\rho_{22}(t)$ and $\rho_{uu}(t)$ and when $\bar{n}\neq 0$. The oscillations of the populations are not as pronounced as in the case of $\rho_{22}(0)=1$. The population $\rho_{ww}(t)$ decays without oscillation. A physical understanding of this behaviour is again provided by the temporal behaviour of the coherence $\rho_{2u}(t)$. From the expression (\ref{e48}) one can see that in the case of $\rho_{ww}(0)=1$ the coherence is different from zero only if $\bar{n}\neq 0$ and it oscillates with a small amplitude $\gamma/\Omega_{u}$. Although the amplitude of the oscillation is small it is, nevertheless, the case where the coherence is produced  by the incoherent pumping induced by the phonon bath.
\begin{figure}[h]
\begin{center}
\begin{tabular}{c}
\includegraphics[width=\columnwidth]{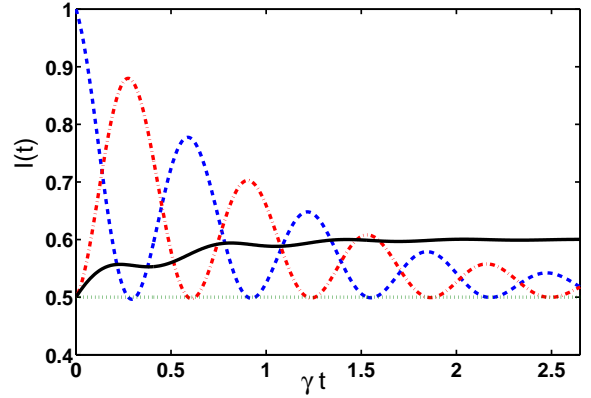}
\end{tabular}
\end{center}
\caption{(Color online) Time evolution of the radiation intensity for $\Omega_{w}=0$, $\Omega_{u}=5\gamma_{0}$, $\gamma_{3}=\gamma_{1}$, $\bar{n}=0$ and different initial conditions: $\rho_{22}(0)=1$ (blue dashed line), $\rho_{uu}(0)=1$ (red dashed-dotted line) and $\rho_{ww}(0)=1$ (green dotted line). The solid black line shows the intensity for the initial state $\rho_{ww}(0)=1$ and $\bar{n}=0.5$. }
\label{fig8}
\end{figure}

Figure~\ref{fig8} shows the radiation intensity $I(t)$, calculated from Eq.~(\ref{e49}), as a function of time for different initial conditions. The coherent exchange of the population between the states $\ket 2$ and $\ket u$ shows up clearly as oscillations in $I(t)$.
We see that even when the evolution starts from the state $\ket w$, the intensity has oscillatory features but only if $\bar{n}\neq 0$. The lack of the oscillations when $\bar{n}=0$ is linked to the fact in this case the initial population remains in the state $\ket w$ for all times.

\section{Effect of spontaneous emission}\label{sec6}

In the preceding sections we have assumed that the QD is exclusively coupled a phonon bath. This assumption is of course an idealization since in reality the phonon modes may not occupy all the modes to which the QD is coupled. In this case, the ordinary spontaneous emission due to the coupling of the QD to vacuum modes can be significant. Therefore, we now include the spontaneous emission, the decay to modes different than that occupied by the phonon modes. The spontaneous emission occurs with rates~$\Gamma_{2}$ and $\Gamma_{3}$, which are the damping rates of the levels $\ket 2$ and $\ket 3$, respectively. In what follows, we illustrate the effect of the spontaneous emission on population inversions between the bare states of the QD and squeezing.

\begin{figure}[h]
\begin{center}
\begin{tabular}{c}
\includegraphics[width=\columnwidth]{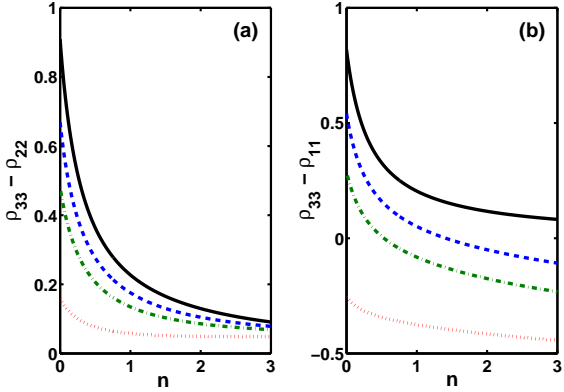}
\end{tabular}
\end{center}
\caption{(Color online) Variation of (a) one-photon $(\rho_{33}(t)-\rho_{22}(t))$ and (b) two-photon $(\rho_{33}(t)-\rho_{11}(t))$ population inversions with the average number of phonons $\bar{n}$ for the case $\Omega_{w}=0$ with $\Omega_{3}=5\gamma_{0}$, $\gamma_{1}=\gamma_{0}$, $\gamma_{3}=10\gamma_{0}$ and different spontaneous emission rates $\Gamma_{2}=\Gamma_{3}\equiv \Gamma$: $\Gamma=0$ (solid black line), $\Gamma =\gamma_{0}$ (dashed blue line), $\Gamma=2\gamma_{0}$ (dashed-dotted green line), $\Gamma=5\gamma_{0}$ (dotted red line).}
\label{fig9}
\end{figure}

To describe the effect of spontaneous emission on the behaviour of the QD, we solve numerically the set of coupled equations of motion (\ref{e22}) with $\Gamma_{2}$ and $\Gamma_{3}$ included. Figure~\ref{fig9} shows the influence of the spontaneous emission on the steady-state population inversions between the bare energy states of the QD. The frame (a) shows the one-photon population inversion between the states $\ket 3$ and $\ket 2$, while frame (b) shows the two-photon population inversion between the states $\ket 3$ and $\ket 1$. We observe that the effect of the spontaneous emission on the one-photon inversion is not dramatic, it decreases with increasing $\Gamma$ but the inversion can still be present even for relatively large spontaneous emission rates. On the other hand, the two-photon inversion decreases rapidly with $\Gamma$ and only at small $\bar{n}$ survives for large spontaneous emission rates. The effect of increasing $\Gamma$ is clearly to decrease the inversion and also to restrict further the range of $\bar{n}$ at which it occurs.

\begin{figure}[h]
\begin{center}
\begin{tabular}{c}
\includegraphics[width=0.9\columnwidth]{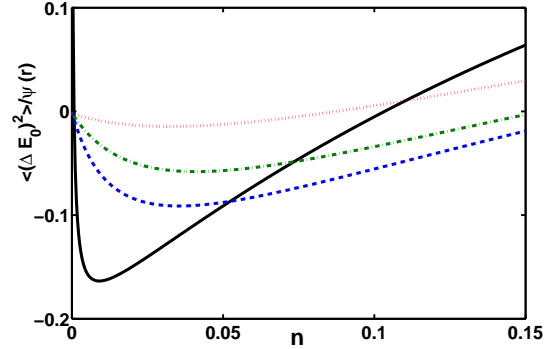}
\end{tabular}
\end{center}
\caption{(Color online) The variance $\langle :(\Delta E_{0})^{2}:\rangle/\psi^{2}(\vec{r}\,)$ as a function of $\bar{n}$ for $\Omega_{w}=\Omega_{u}\, (\Omega_{3}=0.1\gamma_{0})$, $\gamma_{1}=5\gamma_{0}, \gamma_{3}=\gamma_{0}$\, and different spontaneous emission rates $\Gamma_{2}=\Gamma_{3}\equiv \Gamma$: $\Gamma=0$ (solid black line), $\Gamma =0.1\gamma_{0}$ (dashed blue line), $\Gamma=0.2\gamma_{0}$ (dashed-dotted green line), $\Gamma=0.5\gamma_{0}$ (dotted red line).}
\label{fig10}
\end{figure}

Figure~\ref{fig10} shows the effect of the spontaneous emission on the steady-state variance $\langle :(\Delta E_{0})^{2}:\rangle/\psi^{2}(\vec{r}\,)$ for the case of the symmetric driving with $\Omega_{w}=\Omega_{u}$. For $\Gamma=0$ the magnitude of squeezing is large but it occurs in a restricted range of $\bar{n}$. The effect of increasing the spontaneous damping is to decrease the amount of squeezing, but it is interesting that squeezing occurs in less restricted range of $\bar{n}$.

We are therefore led to the conclusion that the effect of the spontaneous emission on the population inversions and squeezing is not dramatic that these features should be observed as long as $\Gamma_{2},\Gamma_{3}<\gamma_{1},\gamma_{3}$, that is, when the spontaneous emission rates are smaller than the rates associated with the coupling of the quantum dot to the phonon bath. In practical terms it would correspond to the situation where the majority of the modes with which the quantum dot interacts is occupied by the phonon modes.

\section{Summary}\label{sec7}

We have studied the effect of a low frequency (acoustic) phonon bath on the dynamics of a quantum dot modelled as a three-level system in a cascade configuration. The phonon bath forms a broadband multimode thermal reservoir to the quantum dot. We have found that the phonon bath couples to the upper transition of the three-level system as an inverted harmonic oscillator that it serves as a linear amplifier to the system, and thereby gives rise to unusual features in the dynamics of the quantum dot. One of these features is the decay of the system to a correlated two-photon state with the population distribution no longer obeying a Boltzmann distribution. We have also found that the radiation field emitted by the quantum dot may exhibit significant squeezing and strong two-photon correlations. These effects are intrinsically connected to a pure two-photon state to which the system decays under the influence of the phonon bath. We have calculated the radiation intensity and have shown that the correlations induced by the phonon bath are manifested in the presence of quantum beats in the time evolution of the radiation intensity. Finally, we have included the ordinary spontaneous emission due to the coupling of the quantum dot to vacuum modes different than that occupied by the phonon modes, and have found that the effect of the spontaneous emission  on the unusual features is not dramatic. The population inversions and squeezing should be observed as long as the spontaneous emission damping rates are smaller than the rates associated with the interaction of the quantum dot with the phonon bath.

\section*{Acknowledgment}

We acknowledge financial support from the National Natural Science Foundation of China (Grant No. 61275123) and the National Basic Research Program of China (Grant No. 2012CB921602).

\end{document}